\documentclass[aps,prl,twocolumn,amssymb,groupedaddress,showpacs,floatfix]{revtex4}
\usepackage{graphicx}
\def\be{\begin{equation}}
\def\ee{\end{equation}}
\def\ba{\begin{eqnarray}}
\def\ea{\end{eqnarray}}
\def\bc{\begin{center}}
\def\ec{\end{center}}

\begin{document}
\title{Magnetotransport in 2D electron systems with a Rashba spin-orbit interaction}

\author{M. V. Cheremisin, A. S. Furman}
\affiliation{A.F.Ioffe Physical-Technical Institute,
St.Petersburg, Russia}
\date{\today}

\begin{abstract}
The beating pattern of Shubnikov-de Haas oscillations in 2D
electron system in the presence of a Rashba zero-field spin
splitting is reproduced. It is shown, taking into account the
Zeeman splitting, that the explicit formulae for the node position
well describes the experimental data. The spin-orbit interaction
strength obtained is found to be magnetic field independent in an
agreement with the basic assumptions of the Rashba model.
\end{abstract}

\pacs{73.20.At,71.43.Qt,72.25.Dc, 73.61.-r}

\maketitle

There has been growing interest in the zero-magnetic-field spin
splitting\cite{Lommer85,Luo88,Das90} of the 2D electron gas
(2DEG), associated with the spin-orbit interaction (SOI) caused by
the structural inversion asymmetry in
heterostructures\cite{Rashba}. Application of a gate voltage
\cite{Nitta97,Engels97} is known to be the most effective method
to control the SOI strength. These 2D systems have been suggested
for application in future spintronics devices, such as spin-based
field-effect transistors\cite{Datta90}, spin-interference
devices\cite{Quan94,Nitta99}, and nonmagnetic spin filters based
on a resonant tunneling structure\cite{Koga02a}. Usually, the
beating-pattern analysis of Shubnikov-de Haas oscillations ( SdHO
) \cite{Das89} and the weak antilocalization method \cite{Koga02b}
are used to determine the SOI strength in 2D systems. However, the
former approach is known to lead to a certain controversy in
determining the zero-field spin splitting $\Delta$. Namely, the
spin splitting deduced from the SdHO beating node position at
finite fields \cite{Das89} is different from that $\Delta$
expected for $B=0$. In the present paper, this discrepancy is
attributed to the contribution of the nonzero Zeeman spin
splitting at finite fields. We support our idea by a rigorous
analysis of the SdHO beating pattern caused by SOI spin splitting.
The beating node positions reported in \cite{Das89} agree well
with those predicted by the theory. Then, we demonstrate that the
SOI strength is independent of the magnetic field.

Let us consider a 2DEG in the x-y plane, subjected to a magnetic
field. In the Landau gauge, the one-electron Hamiltonian including
the Rashba spin-orbit term \cite{Rashba} is given by
\begin{equation}
H=\frac{({\bf p}+e{\bf A})^{2}}{2m}+\frac{\alpha}{\hbar}[{\bf
\sigma}({\bf p}+e{\bf A})]{\bf n}+\frac{g\mu_{B}}{2}({\bf \sigma
B}) \label{hamiltonian}
\end{equation}
where $\bf{p}$ is the 2D momentum; $m$, the effective mass; $g$,
the Zeeman factor; $\mu_{B}$, the Bohr magneton; and, ${\bf n}$,
the unit vector in the z-direction. Then, $\bf{ \sigma}$ is the
Pauli spin matrix; $\bf{B}$, the total magnetic field; and,
$\alpha$, the SOI strength.

It has been shown \cite{Rashba} that the solution to
Eq.(\ref{hamiltonian}) has an explicit form in the case of a
perpendicular magnetic field ${\bf B}=B_{z}=B$. The spectrum for
dimensionless energy $\varepsilon=E/\mu$ ($\mu$ is the Fermi
energy) is given by \cite{Rashba}
\begin{eqnarray}
\varepsilon_{0}=\eta\beta,
\label{spectrum} \\
\varepsilon_{n}^{\pm}=\eta(n \pm \sqrt{\gamma^{2}n+\beta^{2}}), n
\ge 1 \nonumber
\end{eqnarray}
where $\eta=\hbar\omega_{c}/\mu $ is the dimensionless magnetic
field; $\omega_{c}= \frac{eB}{mc}$, the cyclotron frequency;
$\beta=\frac{1}{2}(1-\chi)$, the term containing the Zeeman spin
splitting; $\chi=\frac{g m}{2m_{0}}$, the spin susceptibility; and
$n$, an integer similar to that in the conventional description of
the Landau levels(LL). Then, according to Ref.\cite{Rashba}
$\gamma=\sqrt{\frac{\delta}{\eta}}=\frac{\alpha k_{F}}{\mu
\sqrt{\eta}}$, where $\delta$ is the dimensionless SOI strength
parameter; $\Delta=2 \alpha k_{F}$, the zero-field spin-orbit
splitting at the Fermi energy; and $\hbar k_{F}$, the Fermi
momentum. Usually, the typical Fermi energy $\mu \sim 80 $meV
exceeds the SOI- induced splitting $\Delta \sim 1 $meV ( see
\cite{Das89} ), and, therefore $\delta \ll 1 $. It is noteworthy
that the conventional spin-up(down) energy states associated with
$n$-th LL number correspond to $\varepsilon_{n}^{+}$ and
$\varepsilon_{n+1}^{-}$ states respectively. In the absence of
SOI, Eq.(\ref{spectrum}) reproduces well-known LL energy spectrum.

In contrast to the conventional formalism extensively used to find
the low-B magnetoresistivity, we use the alternative approach
\cite{Kirby73,Cheremisin01a,Cheremisin01b} which allows to resolve
magnetotransport problem in both the SdHO and Integer Quantum Hall
Effect (IQHE) modes. Moreover, this method was successfully used
in a recent paper \cite{Cheremisin05} to reproduce the SdHO
beating structure in the presence of the zero-field valley
splitting (Si-MOSFET 2D system), and in both the crossed- and
tilted- field configurations. Following the argumentation put
forward in Ref.\cite{Cheremisin01b}, well above the classically
strong magnetic field range $\omega_{c}\tau \gg 1$, where $\tau$
is the momentum relaxation time, 2DEG can be assumed
dissipationless in strong quantum limit when the cyclotron energy
$\hbar \omega_{c}$ exceeds both the thermal energy $kT$ and the
energy related to LL-width $\hbar/\tau_{q}$. Here, $\tau_{q}$ is
the quantum relaxation time. Under the above assumptions
$\sigma_{xx}, \rho_{xx} \simeq 0$. Nevertheless, routine dc
measurements yield \cite{Cheremisin01b} the finite
magnetoresistivity associated with a combination of the Peltier
and Seebeck thermoelectric effects. Within the scenario suggested
\cite{Cheremisin01b}, we obtain the above magnetoresistivity in
the form
\begin{equation}
\rho= \rho_{yx}\frac{\alpha_{2D}^2}{L} \label{magnetoresistivity}
\end{equation}
where $\alpha_{2D}$ is the 2DEG thermoelectric power; $\rho
_{yx}^{-1}=Nec/B$, the Hall resistivity; $N=- {\partial \Omega
\overwithdelims()
\partial \mu }_{T}$, the 2D density, $\Omega=-kT \Gamma \sum \limits_{n}\ln
\left(1+\exp \left(\frac{\mu -\varepsilon_{n}}{kT}\right)\right)$,
the thermodynamic potential; $\Gamma=\frac{eB}{hc}$, the
zero-width LL density of states;
$L=\frac{\pi^{2}k_{B}^{2}}{3e^{2}}$, the Lorentz number; $k_{B}$,
the Boltzmann constant. In fact, the 2D thermoelectric power in
strong magnetic fields is a universal quantity \cite{Girvin82},
proportional to the entropy per electron: $\alpha_{2D}
=-{\frac{S}{eN}}$, where $S=-{\partial \Omega \overwithdelims()
\partial T }_{\mu}$ is the entropy. Both $S,N$, and, therefore, $\alpha_{2D},\rho$
are universal functions of the dimensionless temperature
$\xi=\frac{kT}{\mu}$ and the magnetic field $\eta=2/\nu$, where
$\nu=N_{0}/\Gamma$ is the conventional filling factor, and
$N_{0}=\frac{m}{\pi \hbar ^{2}}\mu$ is the zero-field density of
the strongly degenerate 2DEG in the absence of a SOI-induced
splitting.

\begin{figure} \vspace*{0.5cm}
\includegraphics[scale=0.75]{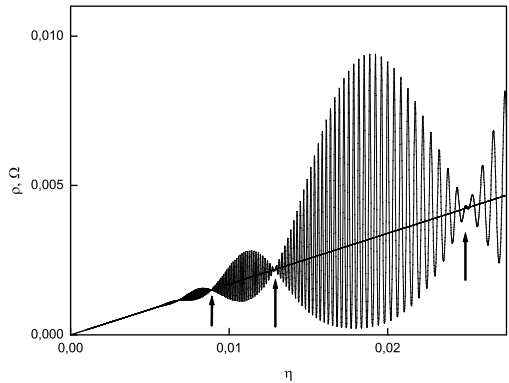}
\caption[]{\label{f2} SdHO beating pattern calculated with the
help of Eq.(\ref{Lifshitz}) at $k=1$ for sample A($x=0.65$)
\cite{Das89}: $N_{0}=1.75*10^{12}$ cm$^{-2}$, $m=0.049m_{0}$,
$g=4$, $\Delta_{0}=2.34$meV, $T=1.6$K. Arrows show the beating
nodes at $j=3,5..$. Zero-field asymptote is represented by dotted
line} \vspace*{-0.5cm}
\end{figure}

Using the Lifshitz-Kosevich formalism and, then, neglecting finite
LL-width( $\hbar/\tau_{q} \rightarrow 0$ ), we derive in Appendix
asymptotic formulae for $\Omega$, and, hence, for
$N,S,\rho_{yx},\rho$, which are valid at low temperatures and weak
magnetic fields $\xi, \eta \ll 1$:
\begin{eqnarray}
N=N_{0}\xi F_{0}(1/\xi )+2\pi \xi N_{0} \sum
\limits_{k=1}^{\infty} \frac{\sin (2\pi k/\eta)}{\sinh
(r_{k})}R(\eta),
\label{Lifshitz} \\
S=S_{0}-2\pi ^{2}\xi k_{B}N_{0} \sum \limits_{k=1}^{\infty }\Phi
(r_{k})\cos(2\pi k/\eta)R(\eta), \nonumber
\end{eqnarray}
where $S_{0}= k_{B}N_{0}(2\xi F_{1}(1/\xi)-F_{0}(1/\xi))$ is the
entropy at $B=0$; $F _{n}(z)$, the Fermi integral; and $\Phi
(z)=\frac{1-z\coth (z)}{z\cdot \sinh (z)}$. At $B=0$ both the
thermopower and 2D density are constants, i.e.
$\alpha_{2D}=\frac{\pi^{2}\xi^{2}}{3}\frac{k_{B}}{e}, N=N_{0}$,
hence the magnetoresistivity is given by zero-field asymptote
$\rho= \frac{h}{e^{2}} \frac{\pi^{2}\xi^{2}\eta}{6}$. According to
Eq.(\ref{Lifshitz}), for actual first-harmonic case( $k=1$ ) the
magnetoresistivity can be viewed as the zero-field background, on
which the rapid SdHO modulated by long-period beatings( see
Fig.\ref{f2} ) are superimposed. It's worthwhile to mention that
at the beat nodes( i.e. when the form-factor at $k=1$ vanishes )
the magnetoresistivity is given by zero-field asymptote. This is
not, however, the case of low temperatures and(or) high magnetic
fields when the high-order terms($k>1$) in Eq.(\ref{Lifshitz}) may
determine the amplitude of magnetoresistivity at the beat nodes.
It turns out that the data reported in \cite{Das89} point to the
above feature.

\begin{figure} \vspace*{0.5cm}
\includegraphics[scale=0.75]{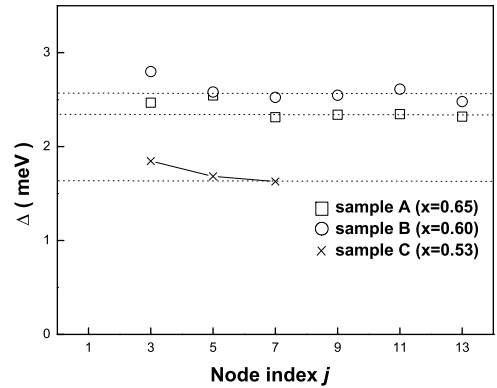} \caption[]{\label{f1}
Zero-field SOI splitting at the Fermi energy vs node index $j$,
deduced from the experimental data \cite{Das89}, with the help of
the node condition specified by Eq.(\ref{node}). Dotted lines(
from top to bottom ) represent the mean values $\Delta_{0}$ for
samples B,A,C respectively.} \vspace*{-0.5cm}
\end{figure}

We now analyze in detail the form-factor $R(\eta)$( see Appendix )
which determines the beating pattern of $S,N$ and, hence, $\rho$.
For the actual first-harmonic case (i.e., $k=1$), the beating
nodes can be observed when $R(\eta)=0$ or
\begin{equation}
\sqrt{\beta^{2}+\frac{\delta}{\eta^{2}}}=\frac{j}{4}, \label{node}
\end{equation}
where we neglect the small quadratic term $\delta^{2}/4 \eta^{2}
\ll \delta/\eta^{2}$ evaluating Eq.(\ref{N+/-}). Then, $j=1,3..$
is the beating node index. We emphasize that the first node cannot
be observed in experiments, performed, for example, in Ref.
\cite{Das89}. Indeed, for real 2D
In$_{x}$Ga$_{1-x}$As/In$_{0.52}$Al$_{0.48}$As system (
$m=0.049m_{0}$, $g \simeq 4$ ) we find $\beta=0.45$, and,
therefore Eq.(\ref{node}) cannot be satisfied for $j=1$. With the
help of Eq.(\ref{node}), we analyze the nodes, reported in
\cite{Das89} for three different samples, and then plot the
dependence of the zero-field SOI splitting at the Fermi energy
$\Delta$ against the node index( see Fig.\ref{f1} ), starting from
$j=3$. For these samples $\Delta$ is nearly constant within the
actual range of the magnetic fields, therefore we obtain the
respective mean values $\Delta_{0}$ denoted in Table
\ref{tab:table1}. Note that the minor deviation of $\Delta$ with
respect to its mean value in high-field limit( low-index nodes )
can be associated with possible magnetic field dependence of the
g-factor. In contrast, the non-parabolicity effects \cite{Hu99}
seem to be irrelevant \cite{Bychkov90} for the actual low-field
case $B<1$T.

We emphasize that the node condition similar to Eq.(\ref{node})
was previously discussed in literature. Following the analysis
done in Ref.\cite{Bychkov90}, the nodes occur when the
spin-orbit-split subbands are shifted one with respect another by
half a period at the Fermi energy. Namely, $1 \simeq
\varepsilon_{n}^{+}=(\varepsilon_{n+s}^{-}+\varepsilon_{n+s+1}^{-})/2$,
where $s=0,1,2...$ corresponds to the node index as $j=1+2s$. For
actual high LL-number case $n \gg 1$ this condition reproduces
Eq.(\ref{node}).

\begin{table}
\caption{\label{tab:table1} Transport data( at 4.2K ) and the
Zero-field spin-orbit splitting at Fermi energy for
In$_{x}$Ga$_{1-x}$As/In$_{0.52}$Al$_{0.48}$As 2D system reported
in Ref.\cite{Das89}}
\begin{ruledtabular}

\begin{tabular}{cccccccc}
Sample($x$)    &$\mu_{0} \times 10^{4}$,cm$^{2}$/Vs &$n \times 10^{12}$cm$^{-2}$ &$\mu$,meV &$\Delta_{0}$,meV \\
\hline   A(0.65)&13.4 &1.75                     &78        &2.34    \\
         B(0.60)&9.5 &1.65                     &74        &2.57    \\
         C(0.53)&6.8 &1.46                     &65        &1.63    \\

\end{tabular}
\end{ruledtabular}
\end{table}

Let us discuss the conventional method \cite{Das89} often used to
extract the zero-field SOI splitting at the Fermi energy.
According to phenomenological arguments put forward by Das et al
\cite{Das89,Das90}, the nodes may occur when $\cos \left(
\pi\frac{\Delta_{tot}}{\hbar\omega_{c}}\right )=0$ or
$\Delta_{tot}=\pm \frac{j}{2} \hbar\omega_{c}$, where the total
spin splitting at the Fermi energy between spin-down
$\varepsilon_{n+1}^{-}$ and spin-up $\varepsilon_{n}^{+}$ states
yields  $\Delta_{tot}=\hbar \omega_{c}-\sqrt{(2\beta\hbar
\omega_{c})^{2}+\Delta^{2}}$. As expected, the total spin
splitting $\Delta_{tot}$ coincides with the zero-field $-\Delta$
and the Zeeman $\chi \hbar \omega_{c}$ spin splitting in low( high
) magnetic field limit respectively. With the help of the
dimensionless units the node condition suggested by Das et al
reads $ \sqrt{\beta^{2}+\frac{\delta}{\eta^{2}}}=\frac{1 \pm
j/2}{2}$, hence, reproduces our result if one selects "$+$" set at
$j \ge 1$. We argue that straightforward procedure ( see
Fig.\ref{f1} ) used to extract $\Delta_{0}$ is, however,
preferable compare to zero-field extrapolation method suggested in
Ref.\cite{Das89,Das90}. Indeed, for low-density samples and(or)
under the temperature enhanced conditions the SdHO amplitude is
suppressed, hence, the low-field nodes become hidden. In this case
the zero-field extrapolation method \cite{Das89} may lead to a
subsequent errors.

Let us now reproduce( see Fig.\ref{f2} ) the SdHO beating pattern
with the nodes occurred in a typical sample( sample A($x=0.65$) in
\cite{Das89}) at $B=0.873;0.46;0.291;0.227;0.183;0.153$T using
Eq.(\ref{Lifshitz}), and previously extracted value of zero-field
SOI splitting $\Delta_{0}=2.34$meV. It's worthwhile to mention
that our results differ with respect to those, which can be
obtained within the conventional formalism in the following: (i)
the low-field quantum interference, classical magnetoresistivity
and 3D substrate parallel resistivity\cite{Das89} background are
excluded within our approach; (ii) in contrast to conventional
SdHO analysis, our method determines the absolute value of
magnetoresistivity, and, moreover, can lead to a gradual
transition \cite{Cheremisin01b} from the SdHO to the IQHE mode.

We argue that the noticeable increase in SdHO amplitude was
observed \cite{Das89} at $B \simeq 0.3$T. This value satisfies the
criterion of the classically strong magnetic field since
$\omega_{c}\tau=4$ while the corresponding cyclotron energy $\hbar
\omega_{c}=8.2$K correlates with that $\sim 9.8$K expected from
T-dependent SdHO-damping factor, i.e. when $2\pi^{2}\xi/\eta \sim
1$. We conclude that the energy associated with LL width $\sim
\hbar/\tau_{q}$ is less or at least equal to the thermal energy.
The above estimates point to validity of zero-width LL model in
this particular case. Nevertheless, since both the temperature and
finite LL width known to suppress the SdHO amplitude in a rather
similar manner, we esteem reasonable to reproduce in Fig.\ref{f2}
the SdHO beating pattern using somewhat higher temperature
$T=1.6$K than that $T=0.5$K reported in \cite{Das89}.

Note that our approach provides a correct number of oscillations
between the adjacent nodes. For example, the number of
oscillations confined between $j=3,5$ nodes (37) correlates with
that (35) observed in \cite{Das89}. A minor point is that our
approach predicts a somewhat lower amplitude of SdHO, compared
with that in the experiment \cite{Das89}. For example, for $j=3$
node( $B=0.873$T in Ref.\cite{Das89} ) we obtain $\rho=0.0035$Ohm.
Actually, one would expect the same order of magnitude for SdHO
amplitude between the proximate nodes ( see $j=3,5$ in
Fig.\ref{f2} ). Our estimation is, however, less than both the
absolute magnetoresistivity $40$Ohm at $B=0.873$T and SdHO
amplitude $\sim 5$Ohm reported in Ref.\cite{Das89}.

In conclusion, we demonstrated the relevance of the
approach\cite{Cheremisin01b} regarding the beating pattern of SdHO
caused by Rashba spin-orbit interactions. Taking into account the
Zeeman splitting, the rigorous analysis of experimental data
\cite{Das89} suggests a B-independent strength of the Rashba SOI.
The above finding is consistent with the general theoretical
assumptions \cite{Rashba}. Our approach can be helpful for
estimation of the SOI strength.

The authors wish to thank Prof. N.Averkiev and Dr. S.Tarasenko for
helpful comments. This study was supported by the Russian
Foundation for Basic Research (grant 03-02-17588) and LSF (
Weizmann Institute).

\section{Appendix}
\label{Lifshitz-Kosevich formalism}

Using the conventional Poisson formulae
\begin{equation}
\sum\limits_{m_{0}}^{\infty }\varphi (n)=\int\limits_{a}^{\infty
}\varphi (n)dn+2\text{Re}\sum\limits_{k=1}^{\infty
}\int\limits_{a}^{\infty }\varphi (n)e^{2\pi ikn}dn,
\label{Poisson}
\end{equation}
where $m_{0}-1<a<m_{0}$, $m_{0}$ the initial value of the
summation, the thermodynamic potential can be represented as the
sum $ \Omega=\Omega_{0}+\Omega_{\sim}$ of the zero-field and
oscillating parts as follows
\begin{eqnarray}
\Omega_{0}=-N_{0} \mu \xi^{2} F_{1}(1/\xi),
\label{LL_Omega1} \\
\Omega_{\sim}=-N_{0} \mu \eta \xi {\text Re}
\sum\limits_{k=1}^{\infty }\int\limits_{0}^{\infty }e^{2\pi ikn}
\ln \left( 1+e^{\frac{1-\varepsilon_{n}^{\pm}}{\xi}} \right )d n,
\nonumber
\end{eqnarray}
where $F_{n}(z)$ is the Fermi integral. For simplicity, we omit
the SOI-induced splitting in zero-field term $\Omega_{0}$ because
$\delta \ll 1 $. The special interest of the present paper is in
the oscillating term $\Omega_{\sim}$ of thermodynamic potential,
which can be strongly affected by spin-orbit-split
subbands($\pm$). After a simple integration by parts, the
oscillating term yields
\begin{equation}
\Omega_{\sim}=N_{0} \mu {\text Re} \sum\limits_{k=1}^{\infty
}\frac{i \eta }{2\pi k } \int \limits_{0}^{\infty } \frac{e^{2\pi
ikn^{\pm}}}{1+e^{\frac{\varepsilon-1}{\xi}}}d\varepsilon
\label{LL_Omega2}
\end{equation}
Using Eq.(\ref{spectrum}), for a certain energy we calculate the
actual high-order LL-like numbers, associated with both the
spin-orbit-split subbands as
\begin{equation}
n^{\pm}(\varepsilon)=\frac{\varepsilon}{\eta}+\frac{\gamma^{2}}{2}
\pm \sqrt{\beta^{2}+ \frac{\gamma^{2}\varepsilon}{\eta}+
\frac{\gamma^{4}}{4}}. \label{N+/-}
\end{equation}
It should be noted that the integrand equation in
Eq.(\ref{LL_Omega2}) is a rapidly oscillating function, which is,
in addition, strongly damped when $\varepsilon
> 1$. The major part of the magnitude of the integral results from the
energy range close to the Fermi energy, when $\varepsilon \sim 1$.
Therefore, $n^{\pm}(\varepsilon)$ can be regarded as smooth
functions of energy, and, hence, can be re-written as $n^{\pm}=
n^{\pm}_{1}+{\partial n^{\pm} \overwithdelims()\partial
\varepsilon}_{1}(\varepsilon-1)$, where we use the designation
$n^{\pm}_{1}=n^{\pm}(1)$ . Under the above assumption, we can
change the lower limit of integration to $-\infty$ and then use
the textbook expression $\int\limits_{-\infty}^{\infty}\frac{e^{i
k y}}{1+e^{y}}dy=\frac{-i\pi}{\sinh(\pi k)}$ for the integral of
the above type. Finally, the thermodynamic potential yields
\begin{equation}
\Omega=\Omega_{0} + N_{0} \mu 2 \pi^{2} \xi^{2}
\sum\limits_{k=1}^{\infty }\frac{\cos(\pi k
(n^{+}_{1}+n^{-}_{1}))R(\eta)}{r_{k}\sinh{r_{k}}}
\label{LL_Omega3}
\end{equation}
where we assume that ${\partial n^{\pm} \overwithdelims()\partial
\varepsilon}_{1}\sim \frac{1}{\eta}$ is valid for the actual case
of high- order Landau levels $n^{\pm} \gg 1$, and $r_{k}=2\pi
^{2}\xi k/\eta$ is a dimensionless parameter related to T-damping
of SdH amplitude. Then, $R(\eta)=\cos(\pi k(n^{+}_{1}-n^{-}_{1}))$
is the form-factor. The oscillatory part of the thermodynamic
potential consists of rapid oscillations $\cos(\pi
k(n^{+}_{1}+n^{-}_{1})) \simeq \cos(2\pi k/\eta)$, on which
long-period beatings governed by the form-factor are superimposed.
As expected, the form-factor is reduced in absence of SOI to a
field-independent constant $R(\eta)=\cos(2\pi k \beta)$, and,
therefore, the beating structure is absent. Using the conventional
thermodynamic definition, we can easily obtain both the entropy
and the density of 2D electrons, specified by
Eq.({\ref{Lifshitz}}).

\end{document}